\newif\ifColor\Colortrue \Colorfalse
\Colortrue

\documentclass[%
 reprint,
superscriptaddress,
]{revtex4-2}
\usepackage[utf8]{inputenc}
\usepackage{amssymb}
\usepackage{amsmath}
\usepackage{amsfonts}
\usepackage{upgreek}
\usepackage{gensymb}
\usepackage{setspace}
\usepackage{booktabs}
\usepackage[english,ngerman]{babel}
\usepackage{mathtools}
\usepackage{mathrsfs}
\usepackage{mhchem}
\usepackage{hyperref}
\usepackage{color}
\definecolor{rltred}{rgb}{0.75,0,0}  
\definecolor{rltgreen}{rgb}{0,0.5,0}
\definecolor{rltblue}{rgb}{0,0,0.75}
\definecolor{rlttublue}{rgb}{0,0.4,0.6}

\usepackage{subcaption}

\usepackage{multirow}
\usepackage{array}
\newcolumntype{P}[1]{>{\centering\arraybackslash}p{#1}}

\usepackage{graphicx}
\graphicspath{{Bilder/}}

\ifColor
  \hypersetup{colorlinks,%
    linkcolor=rlttublue,%
    citecolor=rltgreen,%
    urlcolor=rltblue,%
    linktocpage=true%
  }
\else

  \hypersetup{colorlinks,%
    linkcolor=black,%
    citecolor=black,%
    linktocpage=true%
  }
\fi

\newcommand{\unitS}{{\ensuremath{\,\upmu\text{V}/\text{K}}}}
\newcommand{\unitLambda}{{\ensuremath{\,\text{W}/(\text{m}\text{K})}}}
\newcommand{\unitPF}{{\ensuremath{\,\text{mW}/(\text{m}\text{K}^{2})}}}

\newcommand{\degreeK}{{\ensuremath{\,\text{K}}}}

\newcommand{\eg}{\emph{e.g.\@} }
\newcommand{\ie}{\emph{i.e.\@} }

\begin{document}

\title{Iterative composition optimization in \ce{Fe2VAl}-based thin-film thermoelectrics using single-target sputtering}

\author{A. Riss}
\email{alexander.riss@tuwien.ac.at}  
\affiliation{Institute of Solid State Physics, Technische Universität Wien, 1040 Vienna, Austria}

\author{E. Lasisch}
\affiliation{Institute of Solid State Physics, Technische Universität Wien, 1040 Vienna, Austria}

\author{S. Podbelsek}
\affiliation{Institute of Solid State Physics, Technische Universität Wien, 1040 Vienna, Austria}

\author{K. Schäfer}
\affiliation{Institute of Solid State Physics, Technische Universität Wien, 1040 Vienna, Austria}

\author{M. Parzer}
\affiliation{Institute of Solid State Physics, Technische Universität Wien, 1040 Vienna, Austria}

\author{F. Garmroudi}
\affiliation{Institute of Solid State Physics, Technische Universität Wien, 1040 Vienna, Austria} 

\author{C. Eisenmenger-Sittner}
\affiliation{Institute of Solid State Physics, Technische Universität Wien, 1040 Vienna, Austria}

\author{T. Mori}
\affiliation{International Center for Materials Nanoarchitectonics (WPI-MANA), National Institute for Materials Science, Tsukuba 305-0044, Japan}
\affiliation{University of Tsukuba, Tsukuba 305-8577, Japan}

\author{E. Bauer}
\affiliation{Institute of Solid State Physics, Technische Universität Wien, 1040 Vienna, Austria}

\selectlanguage{english}

\begin{abstract}
Magnetron sputtering inherently exhibits the advantage of dislodging particles from the target in a ratio equivalent to the target stoichiometry. Nevertheless, film compositions often deviate due to element-dependent scattering with the working gas, necessitating the adjustment of the sputtering process. In this work, we explore an unconventional approach of addressing this issue, involving the employment of an off-stoichiometric target. The required composition is obtained through an iterative process, which is demonstrated by \ce{Fe2VAl} and \ce{Fe2V_{0.9}Ti_{0.1}Al} films as case studies. Ultimately, the correct stoichiometry is obtained from \ce{Fe_{1.86}V_{1.15}Al_{0.99}} and \ce{Fe_{1.88}V_{1.02}Ti_{0.13}Al_{0.97}} targets, respectively. Despite the thermoelectric properties falling below expectations, mainly due to imperfect film crystallization, the strategy successfully achieved the desired stoichiometry, enabling accurate film synthesis without the need of advanced sputtering setups.
\end{abstract}
\maketitle

\section{Introduction} \label{sec:introduction}
Thermoelectric materials exhibit the ability to convert heat into electric energy and vice versa. The performance of a material is expressed by the figure of merit $zT = S^2\sigma T / \lambda$, comprising the Seebeck coefficient $S$, the electrical conductivity $\sigma$ and the thermal conductivity $\lambda$. Since the discovery of the Seebeck effect over 200 years ago, thermoelectric research has mainly focused on bulk materials, with current state-of-the-art materials reaching values of $zT>1$ \cite{rogl2015doped, zhao2014ultralow, pei2011high}. Nevertheless, setups utilizing thermoelectric films deposited on various substrates have also demonstrated excellent properties and have proven suitable for thermoelectric applications \cite{venkatasubramanian2001thin, fan2015low, hinterleitner2019thermoelectric, zheng2023harvesting}.

Over time, various different coating techniques have been developed, encompassing solid, liquid and vapor deposition processes \cite{makhlouf2011current}. Among the techniques within vapor deposition, sputtering is a notable example of physical vapor deposition, nowadays commonly employed as magnetron sputtering. The sputter yield, \ie the amount of material removed from the sputter target per incident ion, differs severely between elements \cite{matsunami1983energy}. More specifically, factors such as the sublimation energy, the scattering cross section, the energy and incident angle of the impinging ion and the crystal structure of the target influence the sputter yield \cite{wasa2012sputtering}. Notably, however, a significant advantage of sputtering, in contrast to various other vapor-deposition processes like thermal evaporation, is that after an initial depletion phase, the particle ratio of the ejected atoms aligns with the composition of the target \cite{ohring2001materials, levy2012microelectronic}. Nevertheless, it is important to emphasize that obtaining the desired composition in the resulting film is anything but guaranteed, which presents a serious issue in obtaining reproducible thermoelectric properties. This is attributed to two factors: i) the angular distribution of the ejected atoms is not the same \cite{mahieu2006monte, neidhardt2008experiment, martynenko2012angular, rogov2020angular} and ii) the ejected atoms experience scattering with gas atoms while moving to the substrate, causing deviations from their intended paths \cite{neidhardt2008experiment, shi2008effect, sarhammar2014mechanisms}.

This constitutes a significant challenge, especially in terms of designing high-performance thermoelectric materials, where tiny variations in the composition can have a severe impact on their properties. It can be addressed through various methods. The simplest approach involves multi-target sputtering \cite{boyd2013control, fan2015low, cai2013thermoelectric, lan2020thermoelectric}, employing multiple targets with distinct discharge power to control the composition. These targets may consist of single elements, alloys or compounds. However, this approach requires a suitable sputter chamber equipped with multiple target holders and power supplies.
Another strategy is the use of chips placed on the target to compensate for deficiencies in one or more elements \cite{hiroi2016thermoelectric, fukatani2018improved, machda2023thermoelectric}. These chips can be composed of one or multiple elements and may vary in size. While this allows for a simple adaptation of the composition, the accuracy is not very high compared to other methods.
Furthermore, adjustments of the composition can be achieved by modifying sputter parameters such as the sputter angle \cite{fukatani2018improved, wu2018high, hou1995structure}, power \cite{kim2023homogeneity} or gas pressure \cite{neidhardt2008experiment, hakuraku1998thin, selinder1991resputtering}, among others. This approach works well for two-element alloys, but may yield insufficient results in case of more constituents.

All of these strategies provide the ability to tailor the composition of sputtered films to meet the required stoichiometry, each with its own set of advantages and disadvantages. Several studies on full-Heusler \ce{Fe2VAl}-based films utilized one of these approaches to obtain better stoichiometry and thermoelectric properties \cite{furuta2014fe, hiroi2016thermoelectric, hiroi2016thickness, fukatani2018improved, kurosaki2020crystal, kurosaki2022thermoelectric}. In this work, we present an unconventional approach, which involves the adaptation of the target. By varying the composition of the target, films with the desired ratio of constituents are obtained. While it requires the ability to synthesize adjusted targets, it avoids the necessity of a multi-target sputter chamber and does not hold the same challenges regarding accuracy and reproducibility as simply placing a chip on a target. This approach is successfully demonstrated for \ce{Fe2VAl}-based Heusler compounds, yielding films with the nominal stoichiometry and crystal structure.

\section{Methodology}
\begin{figure}[t!]
\centering
\includegraphics[width=0.48\textwidth]{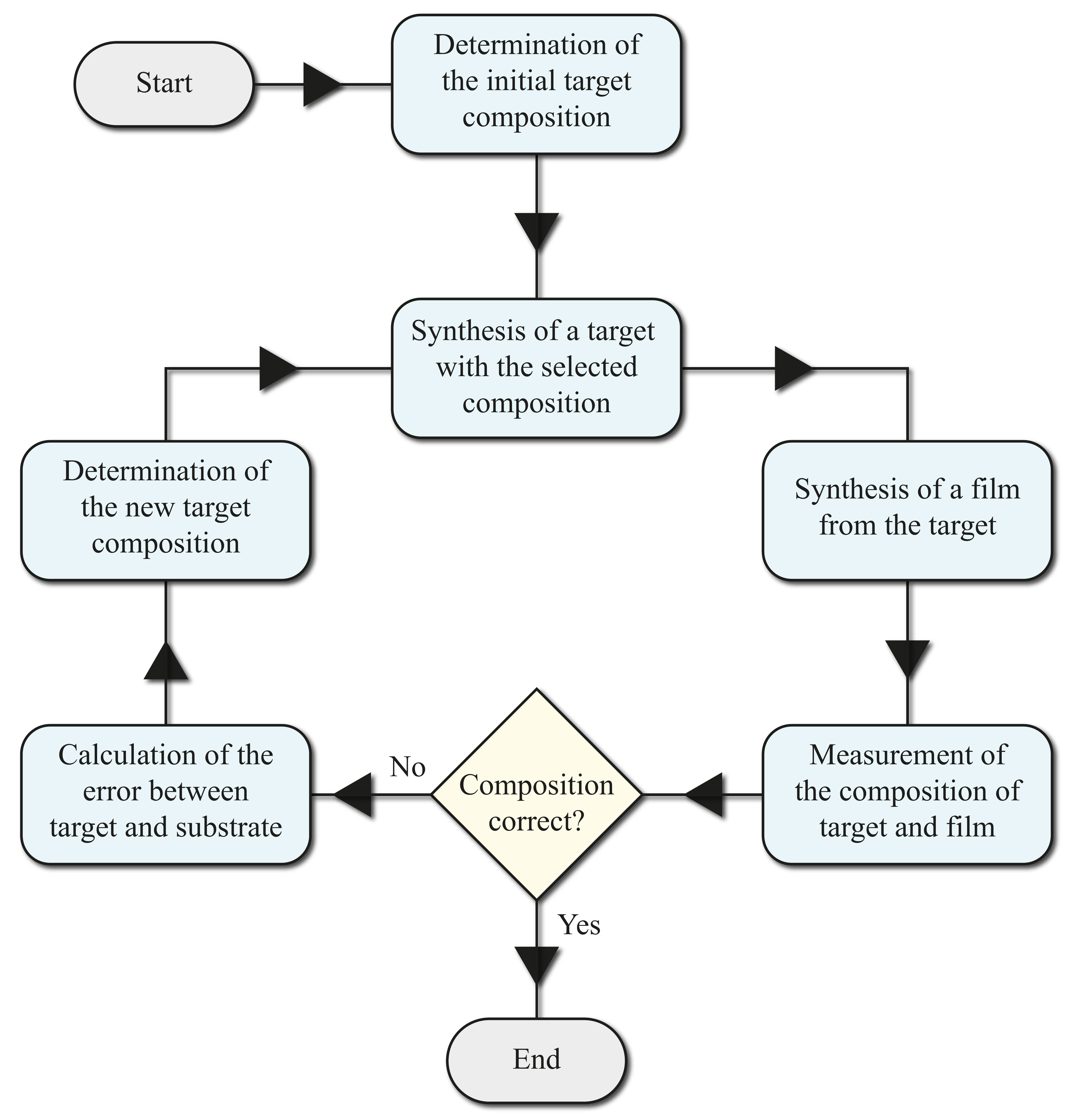}
\caption{Flow chart illustrating the iterative optimization process for refining the film composition through adjustments to the stoichiometry of the target. The iterative process includes the synthesis of both a target and a respective film as well as a measurement of the compositions. The loop repeats until the film's stoichiometry matches the desired composition.}
\label{fig:composition_change_process}
\end{figure}

The stoichiometry of the target is changed according to the deviation of the film's composition. This process is conducted iteratively, starting with the synthesis of a target and the respective film, followed by a composition measurement. If the composition deviates from the desired stoichiometry, an error calculation is performed, followed by the synthesis of another target with an adapted composition. The procedural steps are visually presented in the flow chart shown in Fig.~\ref{fig:composition_change_process}.
The initial target may comprise either the stoichiometric composition, or, when information about deficiencies in one or more constituents is pre-known, an educated guess regarding the off-stoichiometry. 

In this study, the precise composition of the targets has been determined through X-ray fluorescence (XRF) spectroscopy, utilizing the Zetium XRF spectrometer from Panalytical. For thin films, the application of XRF spectroscopy is more complex due to the interdependence of intensity, measured composition and thickness \cite{huang1991thin}. Consequently, the films' composition has been determined using the secondary-electron microscope (SEM) FEI Quanta 250 FEG with an energy-dispersive X-ray (EDX) detector. To minimize measurement errors, the respective target sample has been simultaneously measured as reference. Details of the calculation process are provided in Appendix~\ref{sec:calculation_process}.

Despite the impression that the correct film composition is obtained after the first adoption of the target, a subsequent alteration in the target's composition induces a change in the sputter behavior. This arises \eg from inhomogeneities within the target, resulting from finite solubility of its constituents and the emergence of impurity phases \cite{miyazaki2013thermoelectric}. These impurities may also possess magnetic moments, influencing the magnetic field in the vicinity of the target. Nevertheless, through an iterative process, the deviation of the film's composition from the nominal value can be gradually reduced, as shown below. Each adjustment in the target composition refines the sputter dynamics, ultimately yielding the desired stoichiometry.

The synthesis of the targets was conducted by weighing in high-purity ($> 99.9\,\%$) elements, which were melted together in a homemade high-frequency induction-melting setup. From the obtained ingot, discs with a diameter of $25.4\,\text{mm}$ and a thickness of $2\,\text{mm}$ were cut out. The films were deposited with a direct-current magnetron sputtering setup, employing a discharge power of $10\,\text{W}$ and a working gas pressure of $2\,\text{Pa}$. Furthermore, the target-substrate distance was set to $50\,\text{mm}$, ensuring a uniform thickness distribution.

The crystal structure of all samples was obtained from X-ray diffraction (XRD) measurements, using a Panalytical X'Pert MPDII diffractometer. To minimize reflection peaks from the single-crystalline substrates, an offset of $4^\circ$ was applied.
The thermoelectric properties of both the bulk and film samples were measured utilizing an ULVAC ZEM-3. Furthermore, the thermal conductivity of the bulk material was calculated by determining the density from the lattice parameter obtained from the XRD pattern as well as measuring the thermal diffusivity and heat capacity, employing the Lineis LFA500 Light Flash.

\section{Results}
To elaborate on the practicality and applicability of this approach, \ce{Fe2VAl} and \ce{Fe2V_{0.9}Ti_{0.1}Al} films with the desired composition were synthesized and measured with respect to their structural and thermoelectric properties. In addition, we aimed to synthesize stoichiometric \ce{Fe2TaAl} thin films, which were recently predicted to show superior thermoelectric performance compared to \ce{Fe2VAl}-based systems \cite{bilc2015low, khandy2018case}. However, significant challenges related to the stability of the target material were faced, as elaborated in Appendix~\ref{sec:fe2taal}.

All films had a thickness of $2\,\upmu\text{m}$ and were sputtered on unpolished yttria-stabilized zirconia (YSZ) substrates, ensuring a negligible contribution to the measured Seebeck coefficient and electrical conductivity due to the high electrical resistance \cite{riss2023criteria}.

\subsection{Composition and crystal structure}
\begin{figure}[t!]
\centering
\includegraphics[width=0.48\textwidth]{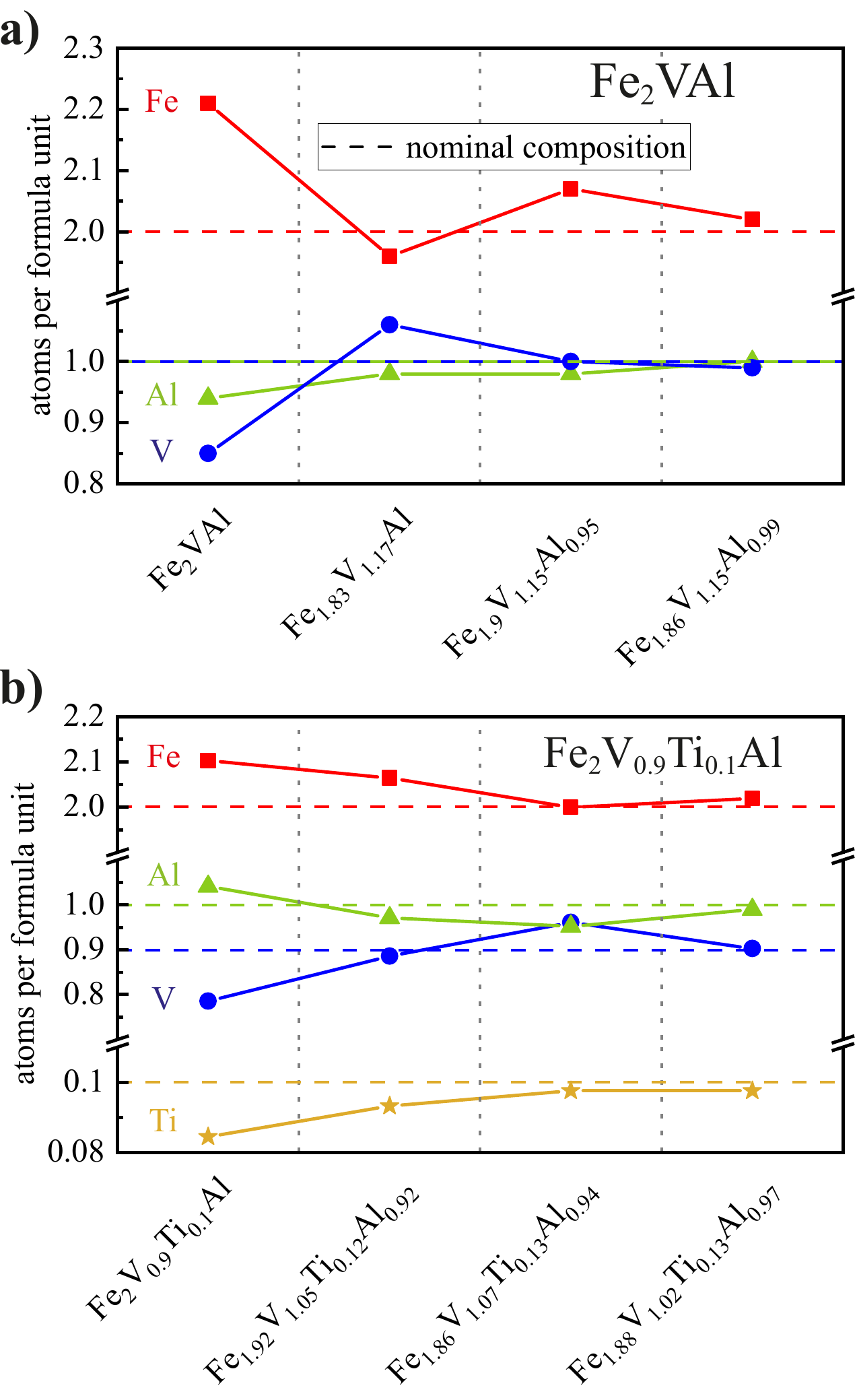}
\caption{Atoms per formula unit of films made from different a) \ce{Fe2VAl}-based and b) \ce{Fe2V_{0.9}Ti_{0.1}Al}-based targets. Nominal stoichiometry is denoted by dashed lines, while solid lines represent the target's composition. The films' composition resulting from the final targets is \ce{Fe_{2.02}V_{0.99}Al} and \ce{Fe_{2.02}V_{0.9}Ti_{0.1}Al_{0.99}}.}
\label{fig:composition}
\end{figure}

Fig.~\ref{fig:composition} presents the composition of \ce{Fe2VAl}- and \ce{Fe2V_{0.9}Ti_{0.1}Al}-based films for each synthesized target, refined according to the above-presented approach.
Films sputtered from the stoichiometric targets exhibited severe off-stoichiometry, \ie \ce{Fe_{2.21}V_{0.85}Al_{0.94}} and \ce{Fe_{2.10}V_{0.79}Ti_{0.08}Al_{1.04}} instead of \ce{Fe2VAl} and \ce{Fe2V_{0.9}Ti_{0.1}Al} were obtained. This underscores the importance for investigating the composition rather than relying on the frequently made assumption that it aligns with the target's stoichiometry. Notably, an excess of \ce{Fe} is evident in both systems, while deficiencies in \ce{V} and \ce{Ti} are observed, similar to previously reported results on \ce{Fe2VAl}-based films \cite{hiroi2016thermoelectric, gao2020significant}. However, it is important to emphasize that the off-stoichiometry can not be a single-element offset but rather represents variations in the concentration of multiple elements.

A total of four iterations (see Fig.~\ref{fig:composition}) were necessary to achieve a satisfactory stoichiometry in the films for both systems. Ultimately, a composition of \ce{Fe_{2.02}V_{0.99}Al} and \ce{Fe_{2.02}V_{0.9}Ti_{0.1}Al_{0.99}} was obtained from \ce{Fe_{1.86}V_{1.15}Al_{0.99}} and \ce{Fe_{1.88}V_{1.02}Ti_{0.13}Al_{0.97}} targets, respectively, closely resembling the desired stoichiometry.

\begin{figure}[t!]
\centering
\includegraphics[width=0.48\textwidth]{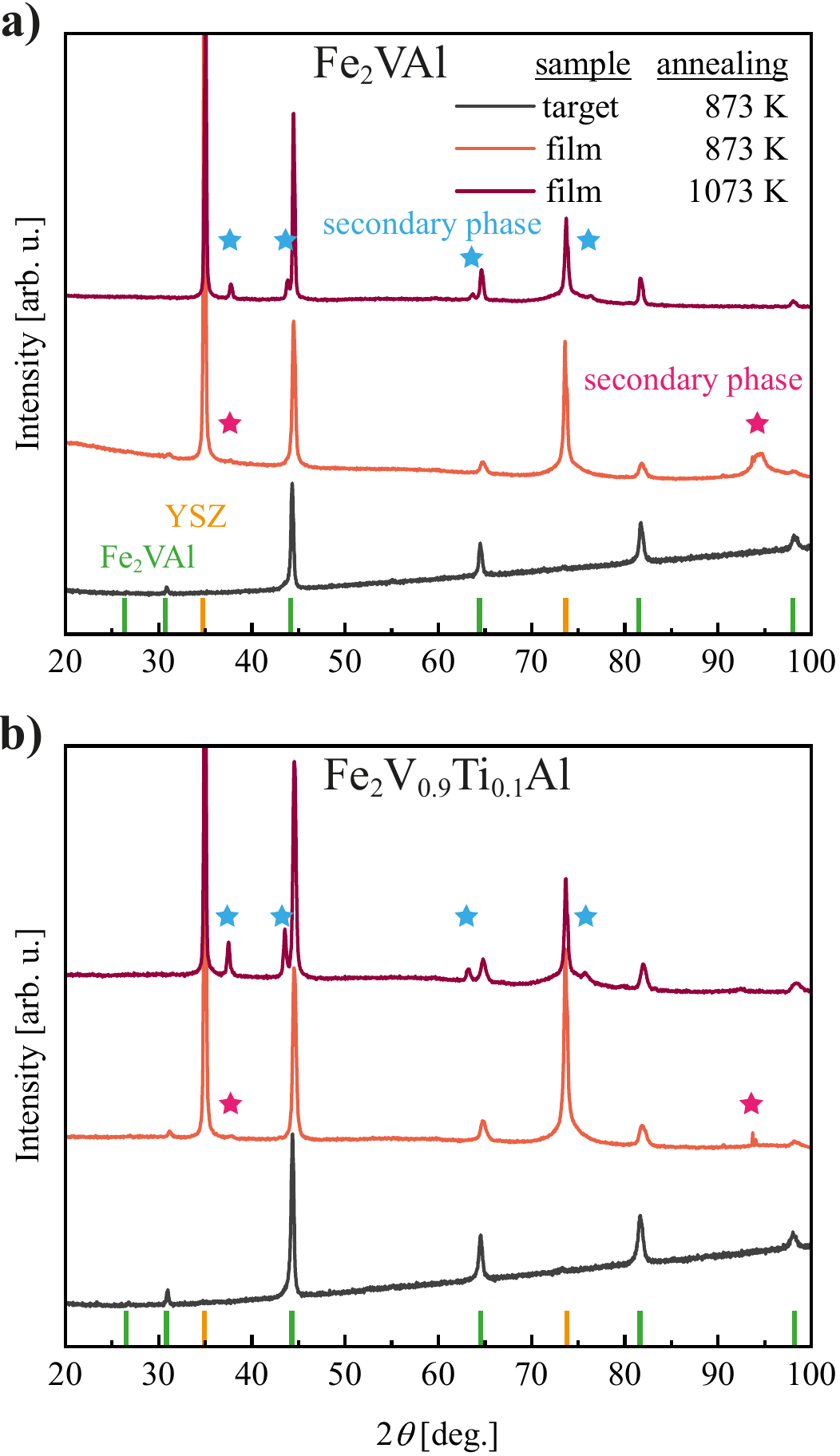}
\caption{X-ray diffraction pattern of a) \ce{Fe2VAl} and b) \ce{Fe2V_{0.9}Ti_{0.1}Al}, featuring the stoichiometric targets (gray line) alongside the films with the closest composition, annealed at $873\degreeK$ and $1073\degreeK$ for 3 days. Peaks from the full-Heusler structure and yttria-stabilized zirconia (YSZ) substrate are highlighted with green and orange marks, respectively. Different impurity phases present in the films are depicted with pink and blue symbols.}
\label{fig:xrd}
\end{figure}

To elucidate the structural characteristics, the X-ray diffraction powder patterns of stoichiometric \ce{Fe2VAl} and \ce{Fe2V_{0.9}Ti_{0.1}Al} bulk specimens, along with diffraction patterns of the final stoichiometric films, are presented in Fig.~\ref{fig:xrd}.
The analysis of the targets reveals the presence of all dominant peaks characteristic of the full-Heusler structure. Of particular significance are the (111) and (200) lines, appearing at $\approx 27\,^\circ$ and $\approx 31\,^\circ$, respectively. These lines serve as indicators of the degree of disorder within the material. In the event of B2 disorder, where a complete disordering of the \ce{V} and \ce{Al} sites occurs, the (111) peak vanishes, while the fully disordered A2 structure lacks both lines \cite{maier2016order}. Although not clearly visible in Fig.~\ref{fig:xrd}, both peaks are present in the targets, suggesting nearly complete ordering.

On the contrary, the films do not exhibit a fully ordered structure. Aside from peaks originating from the YSZ substrate and disorder-independent peaks from the full-Heusler structure, films annealed at $873\degreeK$ reveal only the (200) peak due to the lack of thermal energy necessary for L$2_1$ ordering. Furthermore, upon annealing at $1073\degreeK$, both the (111) and (200) lines are absent. Simultaneously, additional impurity peaks, presumable oxides, form during the annealing process.

\subsection{Thermoelectric properties}
\begin{figure*}[t!]
\centering
\includegraphics[width=0.95\textwidth]{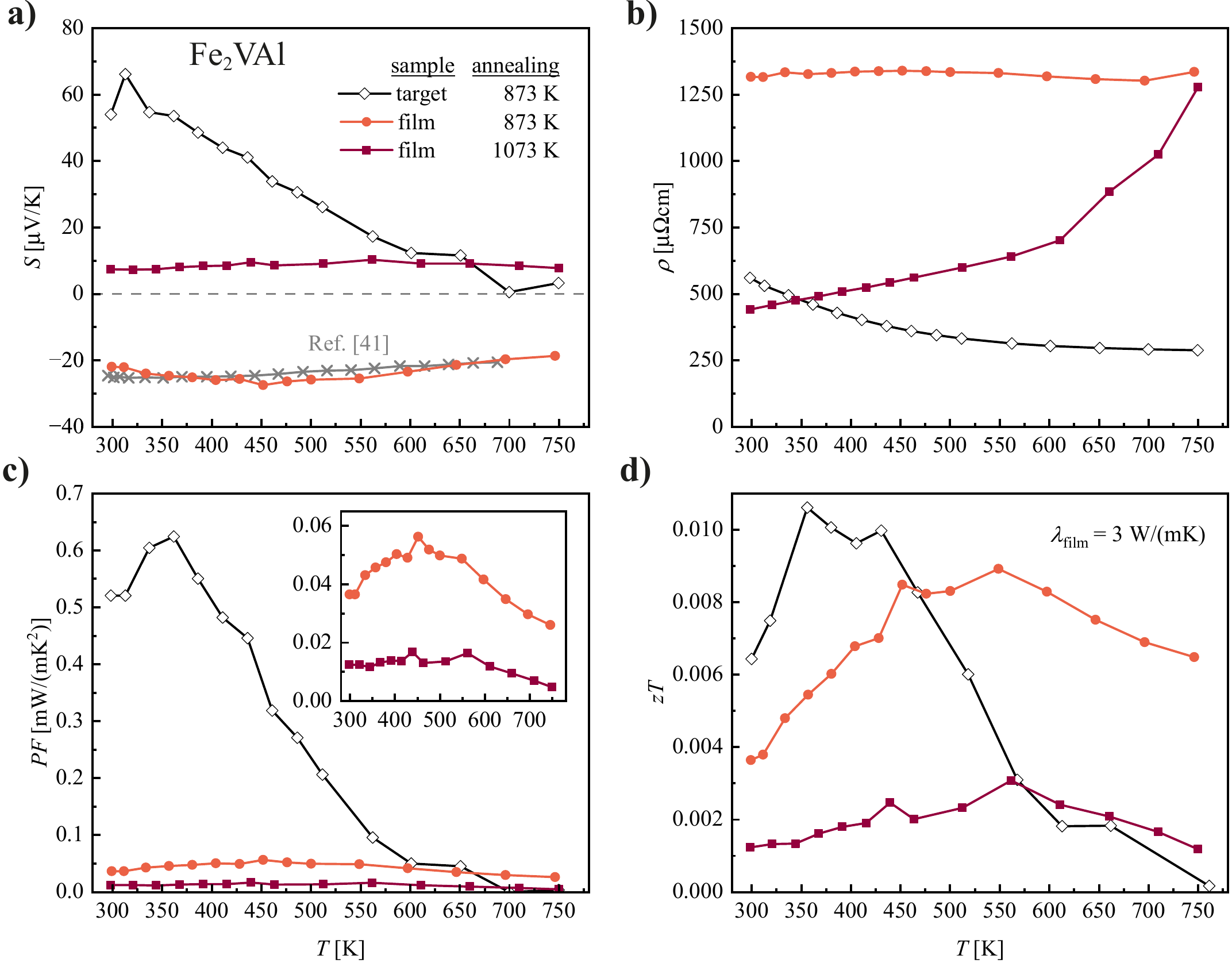}
\caption{a) Seebeck coefficient $S$, b) electrical resistivity $\rho$, c) power factor $PF$ and d) figure of merit $zT$ of the \ce{Fe2VAl} target (black symbols) and the films with the closest stoichiometry after annealing at $873\degreeK$ and $1073\degreeK$ (dark red symbols) as a function of temperature. The $zT$ of the films was calculated assuming a temperature-independent thermal conductivity of $3\unitLambda$, consistent with prior findings \cite{furuta2014fe}. For comparison, the Seebeck coefficient of B2-disordered \ce{Fe2VAl} is included in (a) as gray points, taken from Ref. \cite{garmroudi2022anderson}.}
\label{fig:therm_prop_Fe2VAl}
\end{figure*}

The thermoelectric transport properties of the stoichiometric targets were measured, alongside those of films with the closest composition. The results for \ce{Fe2VAl} are presented in Fig.~\ref{fig:therm_prop_Fe2VAl}.
The electronic structure of \ce{Fe2VAl} features a close-to-zero band gap, resembling a small-gap semiconductor or semimetal \cite{singh1998electronic, berche2020unexpected, hinterleitner2021electronic}. Consequently, disorder or changes in the charge carrier concentration from off-stoichiometry result in significant variations of all thermoelectric quantities \cite{alleno2023optimization}. Consistent with this, an opposite sign of the Seebeck coefficient was previously reported in B2-disordered \ce{Fe2VAl} \cite{garmroudi2022anderson}, aligning with the measured Seebeck coefficient of the B2-disordered film annealed at $873\degreeK$ shown in Fig.~\ref{fig:therm_prop_Fe2VAl}a. The respective electrical resistivity exhibits an increased value and weak temperature dependence, suggesting imperfect formation of the Heusler structure within the film due to limited thermal energy during sputtering and annealing. Furthermore, upon annealing at $1073\degreeK$, resulting in the A2-disordered metallic ground state \cite{alleno2020structure}, the Seebeck coefficient is deteriorated to $< 10\unitS$. In addition, the electrical resistivity decreases due to an increase of the carrier concentration and increasingly metallic ground state, further corroborating the findings from the XRD measurement. 

The measured \ce{Fe2VAl} target exhibits a positive Seebeck coefficient of $\approx 60\unitS$ at room temperature and an electrical resistivity resembling semiconductorlike behavior, consistent with previous reports in literature \cite{nishino1997semiconductorlike, hinterleitner2020stoichiometric, garmroudi2021boosting, parzer2022high}. Given the coinciding composition of target and film, the differences in the temperature-dependent properties and the reduced thermoelectric performance of the film are attributed to finite crystallization and limited degree of order in the structure.

Ultimately, a maximum power factor of \linebreak $5.6\cdot 10^{-2}\unitPF$ at $450\degreeK$ was achieved. Assuming a constant value for the thermal conductivity of $3\unitLambda$ \cite{furuta2014fe}, a figure of merit of $8.2\cdot 10^{-3}$ at $550\degreeK$ was obtained. 

\begin{figure*}[t!]
\centering
\includegraphics[width=0.95\textwidth]{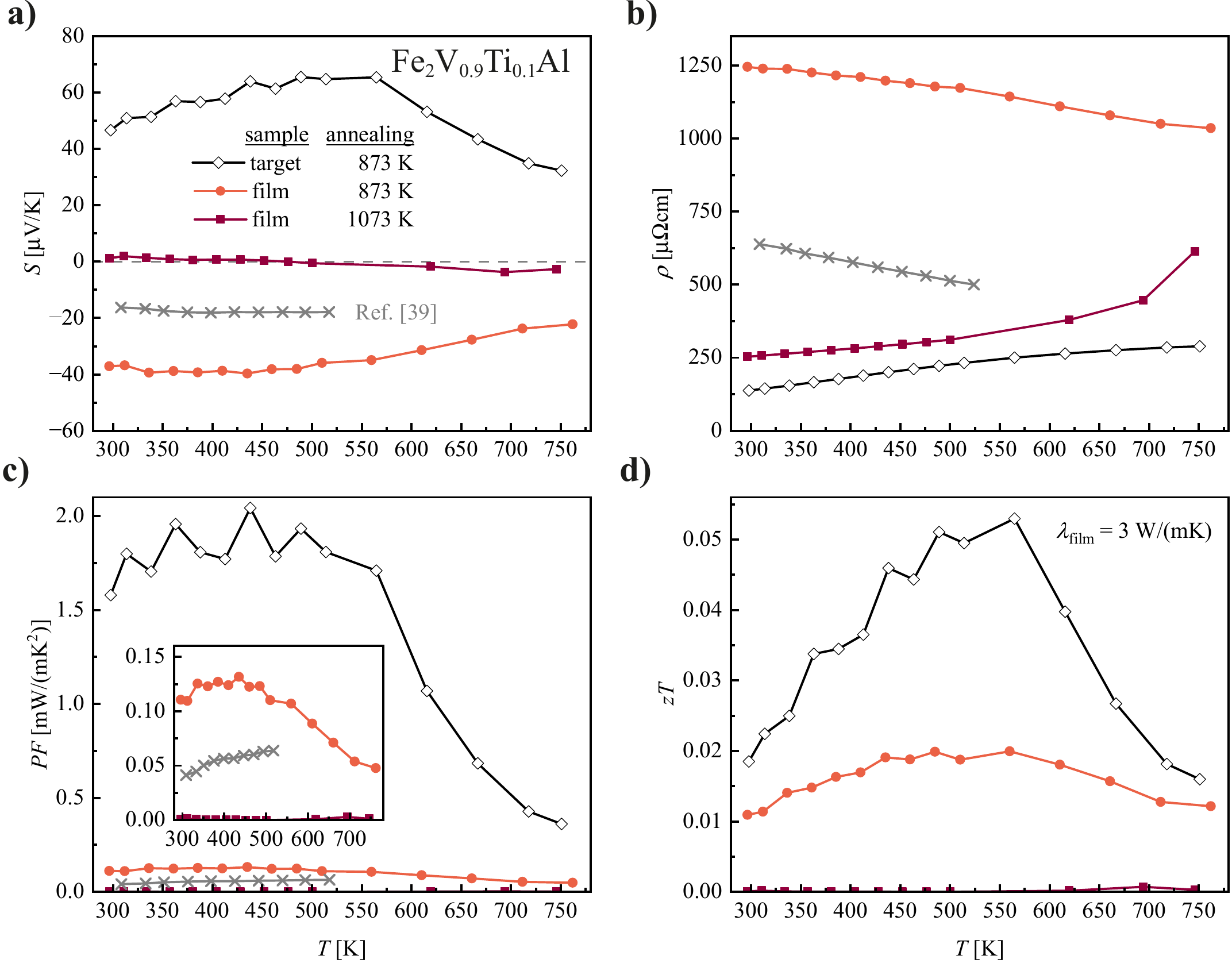}
\caption{a) Seebeck coefficient $S$, b) electrical resistivity $\rho$, c) power factor $PF$ and d) figure of merit $zT$ of the \ce{Fe2V_{0.9}Ti_{0.1}Al} target (black symbols) and the films with the closest stoichiometry after annealing at $873\degreeK$ and $1073\degreeK$ (dark red symbols) as a function of temperature. In addition, the thermoelectric properties of the final \ce{Fe_{1.88}V_{1.02}Ti_{0.13}Al_{0.97}} target are included (blue symbols). The $zT$ of the films was calculated assuming a temperature-independent thermal conductivity of $3\unitLambda$, consistent with prior findings \cite{furuta2014fe}. For comparison, the Seebeck coefficient, electrical resistivity and power factor of off-stoichiometric \ce{Fe2V_{0.9}Ti_{0.1}Al} films are included as gray points, taken from Ref. \cite{gao2020significant}.}
\label{fig:therm_prop_Fe2VTiAl}
\end{figure*}

The thermoelectric properties of both bulk and film \ce{Fe2V_{0.9}Ti_{0.1}Al} are depicted in Fig.~\ref{fig:therm_prop_Fe2VTiAl}.
Similar to the \ce{Fe2VAl} films discussed above, films with a composition resembling \ce{Fe2V_{0.9}Ti_{0.1}Al} exhibit a disorder-induced negative Seebeck coefficient, characteristic for the effect of disorder in \ce{Fe2VAl}-based materials \cite{garmroudi2022anderson}. After annealing at $873\degreeK$, the Seebeck coefficient reaches approximately $-40\unitS$. Concurrently, the resistivity exhibits a small negative temperature dependence, indicative of disorder-dominated transport. Eventually, a broad maximum power factor of $0.12\unitPF$ is achieved between $330$ and $490\degreeK$ (highlighted in the inset of Fig.~\ref{fig:therm_prop_Fe2VTiAl}c). Recently, thermoelectric properties of films sputtered from a stoichiometric \ce{Fe2V_{0.9}Ti_{0.1}Al} target without controlling the composition were reported \cite{gao2020significant}. Comparing these values (depicted as gray data points in Fig.~\ref{fig:therm_prop_Fe2VTiAl}) with our measured results shows the superiority of stoichiometric films and highlights the importance of careful composition tuning for performance optimization. The measured, fully ordered, target exhibits a Seebeck coefficient of $65\unitS$ alongside a small, linearly increasing resistivity, consistent with previously reported results \cite{mikami2013effect}. Moreover, a power factor exceeding $2\unitPF$ and a figure of merit of $0.05$ are achieved. The distinctive differences, compared to the measured films, can be understood by the formation of impurity phases in the latter upon annealing at $1073\degreeK$ and a reduction in the degree of ordering, thereby diminishing the thermoelectric performance. Notably, this further causes the performance to fall below previously reported values of fully ordered \ce{Fe2V_{0.9}Ti_{0.1}Al} films \cite{mikami2010microstructure}.

\section{Conclusion}
In this study, we have proposed and demonstrated a promising approach for tailoring the composition of thermoelectric thin films deposited via sputtering techniques. Traditional methods, such as tuning of deposition parameters and the use of compensatory chips, pose challenges in achieving precise stoichiometry due to various factors influencing sputter dynamics. In contrast, our method involves iteratively adapting the composition of the sputter target to obtain the desired film composition, thus circumventing the need for complex sputter chamber setups and providing greater accuracy compared to chip-based adjustments. 

Films with desired compositions of \ce{Fe2VAl} and \ce{Fe2V_{0.9}Ti_{0.1}Al} were successfully synthesized, demonstrating the applicability of the method and allowing for the gradual refinement of film compositions to closely match the desired stoichiometry, which is crucial for designing high-performance thermoelectric materials. Although the thermoelectric properties of the films were below expectations, structural measurements revealed disorder within the sample, presumably caused by a lack of thermal energy upon annealing, as well as additional secondary impurity phases. Thus, carefully refining the annealing process in future studies will likely yield performances coinciding with the respective bulk material. 
Nevertheless, compared to films sputtered from the stoichiometric target, which exhibit substantially reduced performances, our results show the potential of the here presented method to control the stoichiometry and improve the properties of films deposited by single-target sputtering.

\acknowledgments
Financial support for the research in this paper was granted by the Japan Science and Technology Agency (JST) programs MIRAI, JPMJMI19A1. We acknowledge the X-ray Center at
TU Wien for providing their equipment.
Furthermore, USTEM at TU Wien is acknowledged for providing
the scanning electron microscope to study the composition of the synthesized bulk and film samples.

\appendix
\section{Calculation process of the target composition}
\label{sec:calculation_process}

\begin{figure}[t!]
\centering
\includegraphics[width=0.45\textwidth]{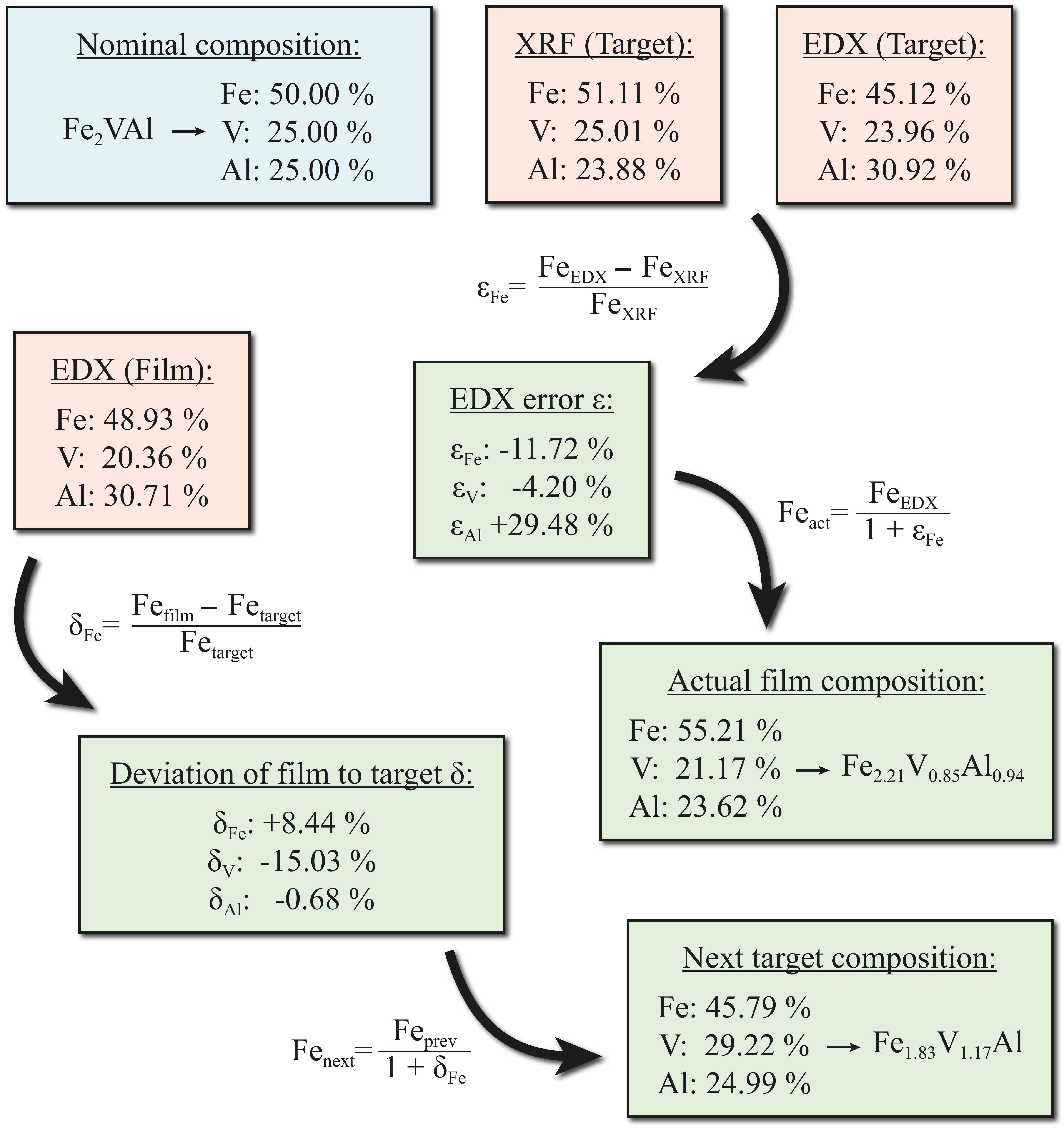}
\caption{Visualization of the sequential process of determining both the film and subsequent target composition. The blue area shows the initially weighed-in, nominal composition of \ce{Fe2VAl}. Measurement data are depicted in orange, while green areas represent calculated values. Equations present next to the arrows highlight the utilized formalism, using \ce{Fe} as an illustrative example.}
\label{fig:calculation_process}
\end{figure}

Fig.~\ref{fig:calculation_process} depicts the process of calculating the film composition as well as the deviation between the target and film stoichiometry.
From the figure it becomes evident that the EDX measurement yields slightly inaccurate results. Assuming that the XRF results depict the correct stoichiometry, the EDX error $\varepsilon$ can be computed for each element. These values are subsequently used to derive the actual composition of the film from the measured values. Furthermore, a comparison of the EDX results from target and films enables the determination of the deviation $\delta$ between the stoichiometries due to different behavior of the constituents during sputtering. Ultimately, $\delta$ is employed to evaluate the required target composition for achieving the correct film stoichiometry.

\section{Synthesis of \ce{Fe2TaAl} films}
\label{sec:fe2taal}

\begin{figure}[t]
\centering
\includegraphics[width=0.45\textwidth]{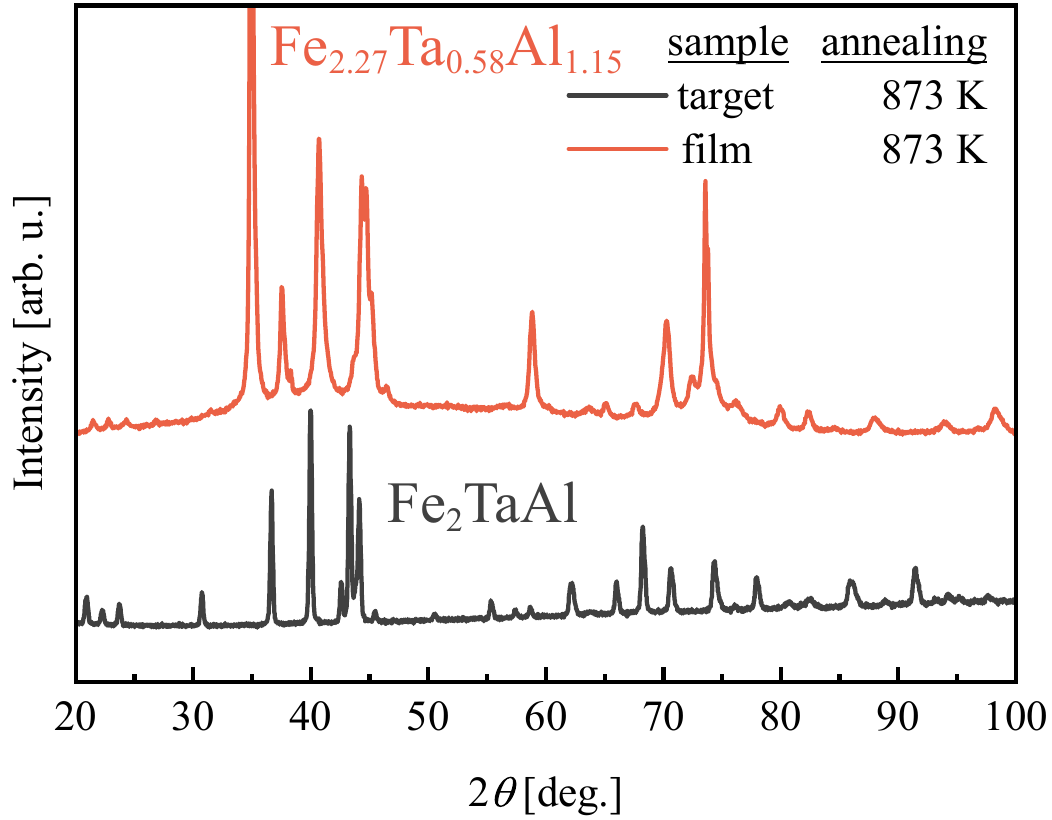}
\caption{X-ray diffraction pattern of the \ce{Fe2TaAl} target (gray line) alongside the respective film with the composition of \ce{Fe_{2.27}Ta_{0.58}Al_{1.15}} (orange line), annealed at $873\degreeK$ for 3 days.}
\label{fig:fe2taal_xrd}
\end{figure}

A stoichiometric \ce{Fe2TaAl} target was prepared, similar to the method described in the main article, by weighing in and melting high-purity \ce{Fe}, \ce{Ta} and \ce{Al}. The films deposited from the target had a composition of \ce{Fe_{2.27}Ta_{0.58}Al_{1.15}}. The X-ray diffraction patterns of both the target and the film, following annealing at $873\degreeK$, are depicted in Fig.~\ref{fig:fe2taal_xrd}.
The abundance of peaks observed in both samples distinctly indicates the presence of multiple phases. Guided by the deviation in the film's composition, an optimized target with a composition of \ce{Fe_{1.81}Ta_{0.87}Al_{1.31}} was synthesized. However, the severe off-stoichiometry caused a complete breaking of the target during the cooling phase subsequent to induction melting due to thermal stress. 

Consequently, small adjustments were made to the composition, leading to the synthesis of \ce{Fe_{1.87}Ta_{0.91}Al_{1.21}}. Despite these refinements, the sample's stability remained compromised, ultimately causing the abandonment of the refinement process for this particular composition.

\newpage

%

\end{document}